\documentclass[11pt]{article}
\usepackage[T1]{fontenc}
\usepackage[utf8]{inputenc}
\usepackage{amsmath,amssymb}
\usepackage[english]{babel}
\usepackage{url}
\usepackage{hyperref}
\usepackage{xcolor}
\newcommand{\quotes}[1]{"#1"}
\usepackage{csquotes}

\usepackage{geometry} 
\begin{document}

\begin{center}
\textbf{\Large{Polynomial analogue of Gandhi's fixed point theorem.\\}}
\end{center}

\begin{center}
\large{Andrey Nechesov}
\end{center}

\begin{center}
Russia, Novosibirsk\\
Sobolev institute of Mathematics\\
\textit{email: nechesov@math.nsc.ru}
\end{center}

\textbf{Abstract:} The problem to be solved in this paper is to construct a general method of proving whether a certain set is p-computable or not. The method is based on a polynomial analogue of the classical Gandhi's fixed point theorem \cite{b_ershov}. The classical Gandhi theorem uses the extension of the predicate with the help of the special operator $\Gamma^{\Omega^*}_{\Phi(x)}$ whose smallest fixed point is the $\Sigma$-set. The work uses a new type of operator - $\Delta_0^p$-operator $\Gamma_{F_{P_1^{+}},...,F_{P_n^{+}}}^{\mathfrak{M}}$, which extends predicates so that the smallest fixed point remains a p-computable set. Moreover, if in the classical Gandhi's fixed point theorem the special $\Sigma$-formula $\Phi(\overline {x})$ is used in the construction of the operator, then in the new operator, instead of a single formula, special generating families of formulas $F_ {P_1 ^ {+}},...,F_{P_n^{+}}$. This work opens up broad prospects for the application of the polynomial analogue of the Gandhi theorem in the construction of new types of terms and formulas \cite{b_cond_terms}, in the construction of new data types and programs of polynomial computational complexity in Turing complete languages. \\


\textbf{Introduction}\\

In both mathematics and programming, we are increasingly confronted with inductively given constructs. These constructs can be both new types of terms and formulas (for example, conditional terms and formula extensions obtained using their \cite{b_gonsvir2019}), and programs or new data types in high-level programming languages that are inductively defined using basic tools. All these inductively generated sets can be viewed as some of the smallest fixed points of a suitable operator. The classical Gandhi theorem allows one to inductively define some abstract set through the special operator $\Gamma^{\Omega^*}_{\Phi(x)}$ \cite{b_ershov} whose smallest fixed point will be $\Sigma$-set. But the $\Sigma$-set is most often not a computable set and moreover not a p-computable set. Therefore, the question arises of how to modify Gandhi's theorem so that the smallest fixed point is a computable or a p-computable set. In this article, we will just talk about the construction of a $\Delta_0^p$-operator whose smallest fixed point is a $p$-computable set, which will allow us to consider many inductive formula definable constructions as some polynomially computable set.\\

\textbf{1.1 P-computability:} Let $\Sigma-$final alphabet and sets $A,B\subseteq\Sigma^{*}$, we say that a function $f:A\to B$ is \textit{p-computable} if there exists a one-tape (multi-tapes) deterministic Turing machine $T$ over the alphabet $\Sigma$ and numbers $C,p\in N$ such that $\forall a\in A$ the value of the function $f(a)$ is calculated on $T$ in at most $C\cdot |a|^p$ steps, where $|a|\geq 1$. A set $A$ is called \textit{p-computable} if such is its characteristic function $\chi_A:\Sigma^{*}\to\{0,1\}$. \textit{The class $P$} of problems solved in polynomial time will often be denoted by $\Delta_0^p$ (accepted notation for the polynomial hierarchy). Therefore, the notation \textit{$\Delta_0^p$-function} for a $p$-computable function and \textit{$\Delta_0^p$-set} for a $p$-computable set will also be used. A partial function $f:A\to B$ is called a \textit{partial $p$-computable} function, if there exists a set $D\subseteq A$ such that $f:D\to B$ is $\Delta_0^p$-function (the Turing machine representing $f$, computes $f(a)$ and stops at the final state $q_0$) and $f(a)$ is undefined if $a\in A\backslash D$ (notation $f(a)\uparrow^p$ or simple $\uparrow^p$), while the Turing machine on the element $a\in A\backslash D$ stops at the final state $q_1$ and finishes its work in $C*|a|^p$ steps. As we can see, a partial $p$-computable function can be extended to a $p$-computable function, but sometimes it is convenient to assume that the value of a $p$-computable function is undefined.\\

Now let $\Sigma_0$ - be some finite alphabet and $\Sigma$\ =\ $\Sigma_0\cup\{\quotes{<}, \quotes{>},\ \quotes{,}\}$ - a new alphabet, obtained by adding new symbols (brackets and comma) to $\Sigma_0$.\\

\textbf{1.2 Word splitting:}\\

\textit{Word splitting} $w\in \Sigma^*$ we will call the following function (that it is a $\Delta_0^p$-function, it will be proved below):
\begin{center}
	$R:\Sigma^*\to \{\Sigma\cup\{\#\}\}^*$, where the symbol $\#$ - new symbol.
\end{center}
such that:
\begin{equation*}
	R(w) =
	\left\{
	\begin{array}{lr}
	w_1\#...\#w_n,\ \text{where }w=<w_1,...,w_n>\text{, every }w_i\in\Sigma^*\text{ and satisfies 1) or 2)}\\
	\uparrow^p, \text{ otherwise }
	\end{array}
	\right.
\end{equation*}	

1) or $w_i\in\Sigma_0^*$\\
2) or $w_i$ - starts with a left bracket and the number of left bracket in the word is equal to the number of right ones, while for any initial subword $\alpha_i$ such that $w_i=\alpha_i.\beta_i$ this is not the case, where the word $w_i$ can be represented as some concatenation of the words $\alpha_i,\beta_i\in\Sigma^*$ and $|\alpha_i|\geq 1$.\\

Inductively define the notion \textit{rank of element} $r(w)$ for $w\in\Sigma^*$:\\
\begin{equation*}
	r(w) =
	\left\{
	\begin{array}{lr}
	0,\ \text{if}\ R(w)\uparrow^p\\
	sup\{r(w_1),...,r(w_k)\}+1, \text{ if } R(w)=w_1\#...\#w_k
	\end{array}
	\right.
\end{equation*}

\textbf{Proposition 1.1} \textit{The splitting is unique.}\\

$\square$ Prove by contradiction. Let there be two different splittings $R(w)=w_1\#...\#w_n$ and $R(w)=l_1\#...\#l_k$ such that $w=<w_1,...,w_n>$ and $w=<l_1,...,l_k>$. But then, by definition, either $w_1,l_1\in\Sigma_0^*$, or $w_1$ and $l_1$ start with a left bracket and the number of right and left bracket for each word is the same. In the first case $w_1$ and $l_1$ are the same. In the second case, $w_1$ is a subword of $l_1$ or $l_1$ is a subword of $w_1$. But then, by definition, no proper subword starting with a left bracket can have an equal number of right and left bracket. Equality of words was also obtained. Further, in a similar way, we show that the remaining $w_i=l_i$ and at the same time $n=k$. $\blacksquare$\\

\textbf{Proposition 1.2} \textit{$R(w)$ is $\Delta_0^p$-function.}\\

$\square$ Consider a Turing machine $T$ with 2 semi-tapes over the alphabet $\Sigma\cup\{0,1,B,\#\}$, where $B$ - is an empty symbol:\\
1) on the first tape we will store the word $w$.\\
2) on the second tape - the number of left and right brackets.\\
\textit{Algorithm of the multi-tape machine:} we write the word $w$ on the first tape, return the head of the 1st tape to the extreme left position and start working with the word. If the first symbol on the first tape is not a left bracket, we finish the work, otherwise we replace it with $B$ symbol and go on to the next steps.\\
1) if the second symbol in the word $w$ is from $\Sigma_0$ then we read up to the first symbol not from $\Sigma_0$. If it's not a comma or a right bracket, then quit.\\
2) if the second symbol is not from $\Sigma_0$ then if it is not a left bracket, then we stop the calculation in the state $q_1$. \\
3) if the second symbol is a left bracket, then further, when reading the left bracket, we add symbol 1 on the 2nd tape and shift to the right, and when reading the right bracket, we erase symbol 1 on the 2nd tape (we moved to the left, put B and stayed in place). If, when reading the right bracket from the first tape, there are no symbols 1 on the second tape, replace the right bracket with $B$ and if there are no other symbols from $\Sigma$ after this right bracket, we stop work in the final state $q_0$, otherwise in the final state $q_1$. \\

\textit{From 1) or 3) we get further:}\\
4) if we meet a comma on the first tape and there are no units on the second tape, then we replace this comma on $\#$ symbol.\\
5) if we come across a right bracket on the first tape in case 1) and no further symbols follow, then we replace this bracket on $B$ and stop the $T$ in the state $q_0$. Otherwise, we stop the computation in the state $q_1$. \\

\textit{Computational complexity $R(w)$:}\\
1) we go by the word on the first tape. Periodically replacing the comma or brackets on $\#$ symbol. The number of such steps does not exceed $2\cdot|w|$.\\
2) on the second tape we write or erase 1 symbols. The number of such added and removed does not exceed $2*|w|$.\\
It turns out that the computational complexity $t(R(w))\leq 4\cdot|w|$. $\blacksquare$\\

\textbf{2.1 Generating formulas and families. False element.}\\

Let $\mathfrak{M}$ is a model of signature $\sigma=\{c_1,...,c_r, f_1^{(m_1)},...,f_s^{(m_s)},R_1^{(p_1)},...,R_t^{(p_t)}\}\cup\{P_1^{(1)},...,P_n^{(1)}\}$ with basic set $M\subseteq\Sigma_0^*$, where $c_l$ - constant symbols ($l\in[1,...,r]$), $f_i$ -- functional symbols ($i\in[1,...,s]$), $R_j$ -- predicate symbols ($j\in[1,...,t]$), $P_k$ -- unary predicate symbols, $k\in[1,...,n]$. $P(\Sigma^*)$ - the set of all subsets of the set $\Sigma^*$. $F_{P_1^+},...,F_{P_n^+}$ - families(\textit{generating families}) positive quantifier-free formulas (which we will call \textit{generating formulas}) of signature $\sigma$ which can include unary predicates $P_1$,...,$P_n$ as $P_j(x_i)$. Moreover, we require that for any free variable $x_i$ in the formula $\varphi_m\in F_{P_k^+}$ there are no occurrences of the form $P_j(x_i)$ and $P_h(x_i)$ for each $x_i$, where $j\neq h$. This property will be called \textit{predicate separability}.\\

The idea is to generate new elements in the form of lists $<a_1,...,a_{n_m}>$, obtained from $(a_1,...,a_{n_m})$ such that $\mathfrak{M}\models\varphi_m(a_1,...,a_{n_m})$, where $a_1,...,a_n\in M$ and then add these elements to the main set of the model:
\begin{center}
$Q_i = \cup_{\varphi_m(x_1,...,x_{k_m})\in F_{P_i^+}}\{<a_1,...,a_{n_m}>\ |\ \mathfrak{M}\models\varphi(a_1,...,a_{n_m})\}$, where $a_1,...,a_{k_m}\in M$
\end{center}
If we want to extend the main set of elements $M$ of the $\mathfrak{M}$ model with this new set of elements $Q_i$, then we need to somehow redefine the functions on these new elements and redefine the truth of the predicates. It is clear that functions on new elements will not be defined, so we will expand the basic set of elements $M$ of the $\mathfrak{M}$ model of signature $\sigma$ with a special $false$-element to $M\cup\{false\}$. Next, we define the semantic meaning of terms and formulas in the $\mathfrak{M}_{false}$ model for all elements from $\Sigma^*\cup\{false\}$ and not only for $M\cup\{false\}$. Since everywhere below only positive quantifier-free formulas with a positive occurrence of unary predicates $P_i$ (in the form of $P_i(x_j)$, for some $x_j$) will appear, then for these formulas on the model 
$\mathfrak{M}_{false}$ we define inductively the values of functions and the truth of predicates, as well as the truth of positive quantifier-free formulas $\varphi_i$, $i\in I$:\\

1) $\mathfrak{M}\models\varphi_i(a_1,...,a_k)\Leftrightarrow\mathfrak{M}_{false}\models\varphi_i(a_1,...,a_k)$, where  $a_1,...,a_k\in M$.\\
2) function value $f_j(a_1,...,a_{n_j})$ equal $false$, if at least one $a_i\in \Sigma^*\cup\{false\}\backslash M$, $j\in[1,...,s]$\\
3) function value $f_j(t_1(\overline{a}),...,t_n(\overline{a}))$ equal $false$, if at least the value of one of the terms $t_j(\overline{a})$ equal $false$.\\
4) formulas of the form $false = t(a_1,...,a_n)$ including $false = false$ we will consider false.\\
5) formulas of the form $a=a$ we will consider true for $a\in M$ and are false otherwise.\\
6) formulas of the form $R_i(t_1(\overline{a}),...,t_{n_i}(\overline{a}))$ we will consider false, if at least of one of the terms $t_j(\overline{a})$ have value $false$.\\ 
7) formulas of the form $P_i(a)$ satisfy 1) and $P_i(false)$ is false.\\ 
8) $\Phi\&\Psi$, $\Phi\vee\Psi$ - retain their standard definitions of truth.\\

\textit{Remark:} so that there are no discrepancies between the $false$-element and false formula, by $false$ - we will mean only the non-standard element introduced above. If we talk about the truth or false of the formula on the model, then if the formula is false, we will denote it as $0$ and if true, then $1$.\\

\textit{Let us denote by $<\mathfrak{M}_{false}, Q>$ - enrichment of the model $\mathfrak{M}_{false}$ such that:}\\
1) $M\cup\{false\}\cup Q$ - new main set.\\
2) all predicates $R_i(t_1,...,t_{n_i})$ remain unchanged if the values of the terms $t_1,...,t_{n_i}$ from $M$ and are false otherwise.\\ 
3) all predicates $P_j(a)$ remain unchanged if $a\in M$ and $P_j(a)$ are false otherwise.\\
4) all functions $f_i(a_1,...,a_n)$ remain unchanged for $a_1,...,a_n\in M$ and have a $false$ value otherwise.\\

Denotation $<\mathfrak{M}_{false}, Q, P_i>$ - it's $<\mathfrak{M}_{false}, Q>$ enrichment, at which the truth set of the predicate $P_i$ is extended to $P_i^{\mathfrak{M}_{false}}\cup Q$.\\

\textbf{2.2. Fixed Points of Monotone Locally Finite Operators.}\\

Let there be a model $\mathfrak{M}_{false}$ of signature $\sigma$ and $Q=(Q_1,...,Q_n)$, where $Q_i\subseteq\Sigma^*$, $i\in[1,...,n]$. Then we introduce the notation:
\begin{center}
$\mathfrak{M}_{false}^{(Q_1,...,Q_n)} = <...<\mathfrak{M}_{false},Q_1,P_1>...,Q_n,P_n>$
\end{center}

Construct an operator:
\begin{center}
 $\Gamma_{F_{P_1^{+}},...,F_{P_n^{+}}}^{\mathfrak{M}}: P(\Sigma^*)\times...\times P(\Sigma^*)\to P(\Sigma^*)\times...\times P(\Sigma^*)\ \ \ $ (1)
\end{center}

which transfer $n$-th sets $(Q_1,...,Q_n)$ to $n$-th sets $(Q'_1,...,Q'_n)$ according to the following rule:

\begin{center}
$Q'_i = Q_i\cup\bigcup_{\varphi_m(x_1,...,x_{k_m})\in F_{P_i^+}}\{<a_1,...,a_{k_m}>\ |\ \ \mathfrak{M}^{(i-1)}_{false}\models\varphi_m(a_1,...,a_{k_m})$, where $\varphi_m\in F_{P_i^+},\ a_1,...,a_{k_m}\in M^{(i-1)} \ \}$
\end{center}

$\mathfrak{M}^{(i-1)}_{false}$ is built on the model $\mathfrak{M}_{false}$ of signature $\sigma$ in the following way:
\begin{center}
$\mathfrak{M}^{(0)}_{false} = \mathfrak{M}^{(Q_1,...,Q_n)}_{false},...,\ \mathfrak{M}^{(i)}_{false} = <\mathfrak{M}^{(i-1)}_{false}, Q'_{i}, P_{i}>$, where $i\in[1,...,n]$.
\end{center}

We fix a partial order $\leq_n$:\\
$(Q_1,...,Q_n)\leq_n(R_1,...,R_n)$, if $Q_i\subseteq R_i$ for all $i\in[1,...,n]$\\

\textbf{Remark:} \textit{operator $\Gamma_{F_{P_1^{+}},...,F_{P_n^{+}}}^{\mathfrak{M}}$ is monotone with respect to the order $\leq_n$, i.e. $(Q_1,...,Q_n)\leq_n(R_1,...,R_n)\Rightarrow \Gamma_{F_{P_1^{+}},...,F_{P_n^{+}}}^{\mathfrak{M}}(Q_1,...,Q_n)\leq_n \Gamma_{F_{P_1^{+}},...,F_{P_n^{+}}}^{\mathfrak{M}}(R_1,...,R_n)$}.\\

\textbf{Remark:} \textit{operator $\Gamma_{F_{P_1^{+}},...,F_{P_n^{+}}}^{\mathfrak{M}}$ possesses the property of a fixed point, i.e. $(Q_1,...,Q_n)\leq_n\Gamma_{F_{P_1^{+}},...,F_{P_n^{+}}}^{\mathfrak{M}}(Q_1,...,Q_n)$}.\\

Associate the operator $\Gamma_{F_{P_1^{+}},...,F_{P_n^{+}}}^{\mathfrak{M}}$ with the sequence: $\Gamma_0,\Gamma_1,...,\Gamma_{t},...$:
\begin{center}
$\Gamma_0=\{\emptyset,...,\emptyset\}\leq_n ... \leq_n\Gamma_{t+1} = \Gamma_{F_{P_1^{+}},...,F_{P_n^{+}}}^{\mathfrak{M}}(\Gamma_{t})\leq_n ... \leq_n \Gamma_{w} = \cup_{k < w}\Gamma_{k}\ \ $(2)
\end{center}

Let us denote by $I_j(\Gamma_k)=Q_j$ (projection onto the $j$-th coordinate).\\

We will say that \textit{operator $\Gamma:P(\Sigma^*)\times...\times P(\Sigma^*)\to P(\Sigma^*)\times...\times P(\Sigma^*)$ - is locally finite}, if for any $X_1,...,X_n\subseteq\Sigma^*$ and any $j\in[1,...,n]$ are done: 
\begin{center}
$I_j(\Gamma(X_1,...,X_n))= \cup_{X'_1\subseteq X_1}...\cup_{X'_n\subseteq X_n}I_j(\Gamma(X'_1,...,X'_n))$
\end{center}
where $X'_1,...,X'_n$ - finite sets.\\

\textbf{Proposition 2.1} \textit{Operator $\Gamma_{F_{P_1^{+}},...,F_{P_n^{+}}}^{\mathfrak{M}}$ is locally finite.}\\

$\square$ Let $X_1,...,X_n\subseteq\Sigma^*$.
Let us show that:
\begin{center}
$I_j(\Gamma_{F_{P_1^{+}},...,F_{P_n^{+}}}^{\mathfrak{M}}(X_1,...,X_n)) = \cup_{X'_1\subseteq X_1}...\cup_{X'_n\subseteq X_n}I_j(\Gamma_{F_{P_1^{+}},...,F_{P_n^{+}}}^{\mathfrak{M}}(X'_1,...,X'_n))\ \ \ $ (3)
\end{center}
where $X'_i$ - finite sets.\\

In one way $\Leftarrow$ inclusion in equality (3) is fulfilled by construction of our operator $\Gamma_{F_{P_1^{+}},...,F_{P_n^{+}}}^{\mathfrak{M}}$, left to show to another.\\

$\Rightarrow$ Let $w\in I_j(\Gamma_{F_{P_1^{+}},...,F_{P_n^{+}}}^{\mathfrak{M}}(X_1,...,X_n))$.
We get that $w$ is a finite list made up of a finite number of elements from $M\cup X_1\cup...\cup X_n$. Mark all the elements involved in constructing w for $w$ from $X_j$ how $C_{j}$, for all $j\in[1,...n]$. Note that all sets $C_{j}$ - are finite and $C_j\subseteq X_j$. Therefore, narrowing our sets $X_i$ to $C_i$ we get, what $w\in I_j(\Gamma_{F_{P_1^{+}},...,F_{P_n^{+}}}^{\mathfrak{M}}(C_1,...,C_n))$. $\blacksquare$\\

\textbf{Proposition 2.2} The smallest fixed point of the operator $\Gamma_{F_{P_1^{+}},...,F_{P_n^{+}}}^{\mathfrak{M}}$ is reached in $w$ steps.\\

$\square$ Claim that the fixed point of the operator $\Gamma_{F_{P_1^{+}},...,F_{P_n^{+}}}^{\mathfrak{M}}$ is reached in $w$ steps follows automatically from the fact that the operator $\Gamma_{F_{P_1^{+}},...,F_{P_n^{+}}}^{\mathfrak{M}}$ is monotone, has the fixed point property and is locally finite. $\blacksquare$ \\

\textbf{2.3 Formulas families $F^*_{P_i^+}$:}\\

Let $\varphi_m(x_1,...,x_{n_m})\in F_{P_i^+}$ - some formula from the generating family, $\epsilon$ - a string of symbols above the alphabet $\{0,1\}$ length $n_m$. Then the formula $\varphi_m^{\epsilon}(x_1,...,x_{n_m})$ - is obtained from $\varphi_m(x_1,...,x_{n_m})$ replacing all occurrences of the form $P_j(x_i)$ on $i$-th symbol in word $\epsilon$, the number of free variables with such a change in the new formula may be less, but we will nevertheless leave their number in the notation for $\varphi_m^{\epsilon}$ as before. 

\begin{center}
$F^*_{P_i^+}=\{\varphi^{\epsilon}_m(x_1,...,x_{n_m})|\ \varphi_m(x_1,...,x_{n_m})\in F_{P_i^+}$, $\epsilon\in\{0,1\}^*$ and $|\epsilon|=n_m$\}
\end{center}
The formula $\varphi_m^{\epsilon}(l_1,...,l_{n_m})$ - is obtained from $\varphi_m^{\epsilon}(x_1,...,x_{n_m})$ by substituting instead of free variables $x_i$ included in the formula $\varphi_m^{\epsilon}$ the corresponding values $l_i$ for all $i\in[1,...,n_m]$. Due to the predicate separability of the formula $\varphi_m$ maximum number of such occurrences in $\varphi_m^{\epsilon}$ may not be more than $n_m$.\\

Define $\Omega=\Sigma\cup\sigma\cup\{0,1\}\cup\{v\}\cup\{\#\}\cup\{\vee,\&\}\cup\{(,)\}$ - a set of symbols such that any formula of the form $\varphi_m(\underline{x_1},...,\underline{x_{n_m}})$, $\varphi_m(l_1,...,l_{n_m})$, $\varphi_m^{\epsilon}(\underline{x_1},...,\underline{x_{n_m}})$, $\varphi_m^{\epsilon}(l_1,...,l_{n_m})$ $\in\Omega^*$, where $l_1,...,l_{n_m}\in\Sigma^*$, $\varphi_m\in F_{P_j^+}$ for some $j\in[1,...,n]$ and $\underline{x_i}$ - encoding the variable $x_i$ with a string of $v$ symbols length $i$.\\

Define \textit{a potentially generating formula} how a formula $\varphi_m(\underline{x_1},...,\underline{x_k})$, potentially generating an element $l$ such that $R(l)=l_1\#...\#l_k$ and  the following holds:
\begin{center}
$\mathfrak{M}^{(l_1,...,,l_k)}_{false}\models \varphi_m^{\epsilon}(l_1,...,,l_k)$
\end{center}
for some signifying $\epsilon$. If for everyone $l\in\Sigma^*$ there is only one potentially generating formula in the family, then we can define a partial function $\gamma_i:\Sigma^*\to\Omega^*$ building by element $l\in\Sigma^*$ its potentially generating formula $\varphi_m(\underline{x_1},...,\underline{x_k})$, if such a formula exists and is undefined otherwise $\gamma_i(l)\uparrow$. In the next chapter we will require from this function to be $p$-computable.\\

\textbf{3.1. $\Delta_0^p$-models and $\Delta_0^p$-operators:}\\

$\mathfrak{M}$ model of finite signature $\sigma$ will be called \textit{$p$-computable model ($\Delta_0^p$-model)}, if all functions are $p$-computable functions, all predicates and the main set are $\Delta_0^p$-sets. Often, instead of writing $\Delta_0^p$-model, will write C-n-$\Delta_0^p$-model, if we want to mark the degree of the polynomial $n$ and the constant $C$. Sometimes, there will be records of the form C-p-$\Delta_0^p$, In the first case $p$ - this is the degree of the polynomial, and in the second $\Delta_0^p$ - this designation for the first level of the polynomial hierarchy. Designation C-p-$\Delta_0^p$-function - we will also apply for functions and C-p-$\Delta_0^p$-set for set. Note that the model $\mathfrak{M}_{false}$ will also C-p-$\Delta_0^p$-computable, if such is the model $\mathfrak{M}$. \\

Operator $\Gamma_{F_{P_1^{+}},...,F_{P_n^{+}}}^{\mathfrak{M}}$ from (1) let's call \textit{$\Delta_0^p$-operator}, if the following 4 properties are holds:\\

\textit{1) $p$-computable model}: $\mathfrak{M}$ is C-p-$\Delta_0^p$-model.\\

\textit{2) predicate separability, quantifier-free and positivity}: each family $F_{P_1^{+}},...,F_{P_n^{+}}$ is either a finite or countable family of formulas, all formulas $\varphi_j\in F_{P_i^{+}}$ are positive, quantifier-free, predicate-separable.\\

\textit{3) uniqueness of the generating formula}: for any two formulas $\varphi_1(x_1,...,x_k)$, $\varphi_2(x_1,...,x_k)$ $\in F_{P_i^{+}}$ with the same number of free variables and for any valuation $\mathfrak{E}:P_{j}(x_i)\to\{0,1\}$, $i\in[1,...,k]$, $j\in[1,...n]$, it is not true, that there exist such signifying $\epsilon_1$ and $\epsilon_2$ from $\mathfrak{E}$, what: 
\begin{center}
$\exists l_1,...,l_k\in M\ $ $\mathfrak{M}_{false}\models\varphi_1^{\epsilon_1}(l_1,...,l_k)\&\varphi_2^{\epsilon_2}(l_1,...,l_k)$
\end{center}

\textit{4) $p$-computability of element}: we also require that all functions $\gamma_i$ to be C-(p-1)-$\Delta_0^p$-functions and
 families $F_{P_i^+}^*$ - C-p-$\Delta_0^p$-families ($t(\mathfrak{M}_{false}\models\varphi^{\epsilon}_m(l_1,...,l_k))\leq C\cdot |l|^p$, for all $\varphi_m\in F_{P_i^+}$ and $l_i\in\Sigma^*\cup\{false\}$), $i\in[1,k]$.\\

\textit{Note} that the $\Delta_0^p$-operator thus defined retains all the original properties: it is monotone, has the fixed point property and is locally finite, and therefore the smallest fixed point of the operator is reached in $w$ steps.\\

We will say that the smallest fixed point $\Gamma_w=(P_1,...,P_n)$ will be $\Delta_0^p$-set, if any $P_i$ -- $\Delta_0^p$-set, where $i\in[1,...,n]$. Let $\gamma_i$ - C-(p-1)-$\Delta_0^p$ - from condition 4) for $\Delta_0^p$-operator $\Gamma_{F_{P_1^{+}},...,F_{P_n^{+}}}^{\mathfrak{M}}$ and $\varphi_m(x_1,...,x_k)\in F_{P_i^+}$ - potential generating formula for $l$, where $R(l)=l_1\#...\#l_k$, all $l_1,...,l_k\in\Sigma^*$. Then the following lemma is true for any signifying $\varphi^{\epsilon}_m(x_1,...,x_k)$:\\

\textbf{Lemma 3.1} $\varphi^{\epsilon}_m(\underline{x_1},...,\underline{x_k})$ is built according to the formula $\varphi_m(\underline{x_1},...,\underline{x_k})$ and by signifying $\epsilon$ for a time not exceeding $12\cdot C\cdot|l|^{p-1}$.\\

$\square$ Consider a Turing machine $T$ over $\Omega$ alphabet consisting of 5 semi-tapes:\\
1) the first tape - the formula is written out $\varphi_m(\underline{x_1},...,\underline{x_k})$, the length of which does not exceed $C\cdot|l|^{p-1}$.\\
2) second tape - written out the word $\epsilon$ of length $k$.\\
3) third tape for variables.\\
4) the fourth tape will remember the last position of the head of the first tape.\\
5) the fifth tape - to build a new formula $\varphi_m^{\epsilon}(\underline{x_1},...,\underline{x_k})$.\\

Let the formula $\varphi_m(\underline{x_1},...,\underline{x_k})$ have already been written out on the first tape and the second tape contains the word $\epsilon$. The machine starts to work in the extreme left position and reads the formula from the first tape. As soon as the T reached the word of the view $P_j(\underline{x_i})$, T begins to read this word and in parallel write out symbols 1 on the 4th tape of the head position and symbols 1 for each symbol $v$ in this word on the third tape, parallel moving the head on the tape containing the word $\epsilon$ with a unit delay. When all the symbols $v...v$ ($\underline{x_i}$) will be read, the head on the tape $2$ will survey symbol $\epsilon_{i_1}$ which must be substituted for the word $P_j(\underline{x_i})$. Since the position of the head of the first tape is recorded on the 4th tape, we begin to overwrite the word $P_j(\underline{x_i})$ on symbols $\#$ on the first tape and reduce in parallel the number of symbols 1 on the 4th. As soon as there are no 1 symbols left on the 4th tape, then write the $\epsilon_{i_1}$ symbol to the first tape. Next, we return the heads of tapes 2, 3, 4 to the extreme left position and continue to sequentially find and replace the remaining occurrences of the view $P_j(\underline{x_r})$ on the first tape and replace them with symbols $\#$ and $\epsilon_{r}$. After all the replacements, we must return the head of the first tape to the extreme left position and start copying the formula of the first tape to the 5th one, while skipping the symbols $\#$.\\

\textit{The total running time of the above algorithm:}\\
1) On the first tape, we work with the formula $\varphi_m(\underline{x_1},...,\underline{x_{n_m}})$, which includes words like $P_j(\underline{x_i})$. In total, we need to go no more than 3 lengths $\varphi_m(\underline{x_1},...,\underline{x_{n_m}})$: read word $P_j(\underline{x_i})$ at the beginning, then replace each of its symbol with $\#$ and put either 0 or 1 instead of the first symbol, etc.\\
2) On the second tape, the machine does not change the word in any way $\epsilon$, simply reads it in parallel with the $v$ symbols from the first tape and periodically returns the head to the leftmost position. The total number of shifts to the right of the head of the 2nd tape does not exceed the length of a word on the first tape. The same goes for the number of head shifts to the left. Therefore, the steps on this tape will be done no more than $2\cdot C\cdot|l|^{p-1}$.\\
3) On the third tape, the last monitored variable is written out. The number of head shifts on this tape to the right and to the left does not exceed $2\cdot C\cdot|l|^{p-1}$.\\
4) For the 4th tape - similar $2\cdot C\cdot|l|^{p-1}$.\\
5) to copy the final word from the first tape to the 5th, taking into account the preliminary setting of the head of the first tape to the extreme left position, it will also take no more than $2\cdot C\cdot|l|^{p-1}$. $\blacksquare$\\

Let $\varphi_m(x_1,...,x_k)\in F_{P_i^+}$ potentially generative formula for an element $l$.\\

\textbf{Lemma 3.2} $\varphi_m^{\epsilon}(l_1,...,l_k)$ built by word $l$ and by the formula $\varphi_m^{\epsilon}(\underline{x_1},...,\underline{x_k})$ for a time not exceeding $8\cdot C\cdot|l|^p$.\\

$\square$ Consider a Turing machine $T$ over alphabet $\Omega$ that also consists of 5 semitapes:\\
1) the first tape - the formula $\varphi_m^{\epsilon}(\underline{x_1},...,\underline{x_k})$ is written out, the length of which does not exceed $C\cdot|l|^{p-1}$.\\
2) second tape - written out the word $R(l)=l_1\#...\#l_k$ length not exceeding $|l|$.\\
3) third tape for elements from $l$.\\
4) the 4th tape will remember the last position of the head of the first tape.\\
5) the 5th tape - to build a new formula $\varphi_m^{\epsilon}(l_1,...,l_k)$.\\

The machine starts working with the formula of the first tape, if necessary, simultaneously copying the result to the 5-th tape. If the machine on the first tape reads a symbol that is not $v$, then $T$ copies it to 5-th tape. If we come to the symbol $v$ on the first tape, then we start the process of finding the corresponding $l_i$ for replacement. When reading the $i$-th symbol $v$ in a row from the first tape, we write on the 4th tape $i$-th 1 symbol. Further, as on the first tape a symbol different from $v$ goes, we overwrite symbols 1 on B symbols from the 4th tape, and in parallel with the second tape (where the word $l_1\#...\#l_k$ is written out), with a single delay, we transfer the head to the next symbol $\#$ that comes before the corresponding $l_i$.  When reading all 1 symbols in a row from the 4th tape, the machine will write the corresponding $l_s$ word to the third tape. Now we need to copy the word $l_s$ to the 5th tape and, after copying, clear tapes 3 and 4, moving the head to the extreme left position. By repeating this algorithm on the 5th tape, the word $\varphi_m^{\epsilon}(l_1,...,l_k)$ will eventually be written.\\

Calculate the total operating time of such a machine T:\\

1) the machine $T$ reads a word from the first tape or just stands and waits for further reading. The number of movements to the right does not exceed $C\cdot|l|^{p-1}$\\
2) on the second tape, the head moves both to the right and to the left, but again only reading. Therefore, the number of ticks does not exceed $C\cdot|l|^{p-1}\times 2\cdot|l|\leq 2\cdot C\cdot|l|^p$.\\
3) the third tape is a temporary storage of the word $l_i$, for some $i$. The number of measures also does not exceed $2\cdot C\cdot|l|^p$.\\
4) 4th tape - the number of shifts does not exceed $2\cdot C\cdot|l|^{p-1}$.\\
5) 5th tape - no more $C\cdot |l|^{p-1}$. $\blacksquare$ steps.\\ 

\textbf{3.2 A polynomial analogue of Gandhi's theorem:}\\

\textbf{Theorem 3.1:} \textit{( polynomial analogue of Gandhi's theorem)}\\
Smallest fixed point $\Gamma_w$ (2) of  $\Delta_0^p$-operator $\Gamma_{F_{P_1^{+}},...,F_{P_n^{+}}}^{\mathfrak{M}}$ is $\Delta_0^p$-set.\\

$\square$ The main idea of the proof is to show that the time for checking the truth of the formula $P_i(l)$ on $\mathfrak{M}^{\Gamma_w}_{false}$ does not exceed the time $k\cdot C\cdot r(l)\cdot|l|^p$, where $k$ and $C$ - are fixed constants, and $r(l)$ is the rank of the element $l$ and $r(l)\geq 1$, $i\in[1,...,n]$. Since the rank $r(l)<|l|$, we get that for any $l$ the complexity does not exceed $k\cdot C\cdot |l|^{p+1}$.\\

Without loss of generality, we show this for $P_1(l)$, assuming in the induction step that this estimate is true for all $P_i(l_j)$, where $r(l_j)<r(l)$ and $i\in[1,...,n]$.\\
Induction by complexity $r(l)$ show that $t(P_1(l))\leq 25\cdot C\cdot r(l)\cdot|l|^p$, where the constant $C$ is the maximum for all constants that participate in the complexity constraint for the splitting function $R(l)$, for functions $\gamma_i$ and for the algorithm for checking the truth of the formula $\varphi_m^{\epsilon}(l_1,...,l_{n_m})$, the maximum degree is $p$.\\

\textit{Induction base $r(l)=1$:}\\
Case 1: $\gamma_i(l)\uparrow^p$, then the formula $P_1(l)$ is false.\\
Case2: $\gamma_i(l) = \varphi_m(x_1,...,x_k)$
Then $R(l)=l_1\#...\#l_k$ and all elements of $l_i$ (where $i\in[1,...,k]$) are either elements of the base set $M$ or elements from $\Sigma^*$ for which $R(l_i)\uparrow^p$. Given that all $P_i(l_j)$ are false on $\mathfrak{M}^{\Gamma_w}_{false}$, we can replace $\epsilon = 0...0$ in the potentially generating formula $\varphi_m$ for element $l$ and get:
\begin{center} 
$\mathfrak{M}^{\Gamma_w}_{false}\models P_1(l)\Leftrightarrow\mathfrak{M}^{\Gamma_w}_{false}\models\varphi_m(l_1,...,l_k)\Leftrightarrow\mathfrak{M}_{false}^{\Gamma_w}\models\varphi_m^{\epsilon}(l_1,...,l_k)\Leftrightarrow\mathfrak{M}_{false}\models\varphi_m^{\epsilon}(l_1,...,l_k)$
\end{center}
 The time required to construct a potentially generating formula $\varphi_m(\underline{x_1},...,\underline{x_k})$ using $l$ does not exceed $C\cdot|l|^{p-1}$. Next, we build $\varphi^{\epsilon}_m(\underline{x_1},...,\underline{x_k})$ and $\varphi^{\epsilon}_m(\underline{l_1},...,\underline{l_k})$. The time required for this does not exceed $12*C*|l|^{p-1}$ and $8*C*|l|^p$ (Lemmas 3.1 and 3.2). And the verification of the truth of the last formula for $\mathfrak{M}_{false}$ does not exceed $C*|l|^p$. Summing everything up, we get that the verification time does not exceed $25*C*r(l)*|l|^p$.\\

\textit{The induction step: let for $r(l)=s$ our assumption is true, we will show for $s+1$:}\\

Case 1: $\gamma_i(l)\uparrow^p$, then the formula $P_1(l)$ is false. We get that 
\begin{center}
$t(P_1(l))\leq C\cdot|l|^p\leq 25\cdot C\cdot r(l)\cdot|l|^p$
\end{center}
Case 2: $\gamma_i(l) = \varphi_m(x_1,...,x_k)$
\begin{center}
$\mathfrak{M}^{\Gamma_w}_{false}\models P_1(l)\Leftrightarrow\mathfrak{M}^{\Gamma_w}_{false}\models\varphi_m(l_1,...,l_k)\mathfrak{M}_{false}^{\Gamma_w}\models\varphi_m^{\epsilon}(l_1,...,l_k)\Leftrightarrow\mathfrak{M}_{false}\models\varphi_m^{\epsilon}(l_1,...,l_k)$ 
\end{center}
where $\epsilon$ string of symbols $\epsilon_{i}$ such that $\epsilon_{i}=1$, if formula $P_j(l_i)$ is true on $\mathfrak{M}^{\Gamma_w}_{false}$ model and 0 otherwise.\\

Let's calculate the time spent on all transitions:\\

1) constructing a potentially generating formula $\varphi_m(x_1,...,x_k)$ using $l$ in time $C\cdot|l|^{p-1}$\\

2) determine the truth of all predicates $P_{i_1}(l_1),...,P_{i_k}(l_k)$, which are included in the formula. By the induction hypothesis, we obtain: 
\begin{center}
$\sum_{j=1}^{k} t(P_{i_j}(l_j))\leq \sum_{j=1}^{k} 25\cdot C\cdot r(l_j)\cdot|l_j|^p\leq 25\cdot C\cdot (r(l)-1)\cdot|l|^p$.
\end{center}

3) further, we fix the signifying $\epsilon:P_{j_i}(x_i)\to\{0,1\}$ considering whether the predicate $P_{j_i}(l_i)$ is true or false, if the formula does not include any of the predicates $P_{j_i}$ for the variable $x_i$, then we determine the truth for $P_1(x_i)$ by default.\\

4) By the formula $\varphi_{m}(x_1,...,x_k)$ and by the signifying $\epsilon$ we construct $\varphi_{m}^{\epsilon}(x_1,...,x_k)$. The time required for this does not exceed $12\cdot|l|^{p-1}\leq12\cdot|l|^{p-1}$.\\

5) By the formula $\varphi_{m}^{\epsilon}(x_1,...,x_k)$ and by $l$ we construct $\varphi_{m}^{\epsilon}(l_1,...,l_k)$. The time required for this does not exceed $8\cdot C\cdot|l|^p$.\\

Next, we determine the truth of the formula $\varphi_{m}^{\epsilon}(l_1,...,l_k)$ on $\mathfrak{M}_{false}$ in the time $C\cdot|l|^p$.\\

If we sum up all the time of calculations, then we get the following:
\begin{center}
$t(P_1(l))\leq \sum_{i=1}^k(25\cdot C\cdot r(l_i)\cdot|l_i|^p)+25\cdot C\cdot|l|^p\leq$\\
$\leq 25\cdot C\cdot (r(l)-1)\cdot|l|^p+25\cdot C\cdot|l|^p\leq 25\cdot C\cdot r(l)\cdot|l|^p$
\end{center}

We have shown that for any element $l$ of rank $r(l)$ in time $25\cdot C\cdot r(l)\cdot|l|^p$ we determine whether it belongs to the predicate $P_1$. But since $r(l)$ is always less than $|l|$, we can write the following:
\begin{center}
$t(P_1(l))\leq 25\cdot C\cdot r(l)\cdot|l|^p\leq 25\cdot C\cdot|l|^{p+1}$.
\end{center}

$\blacksquare$\\

\textbf{3.3. Corollaries and applications}\\

For the C-p-$\Delta_0^p$-model $\mathfrak{M}$ as an application of the polynomial analogue of Gandhi's theorem, we present several corollaries. Some of these corollaries have already been proven earlier by other authors using other methods, some are cited for the first time.\\ 

Let the model $\mathfrak{M}$ have a one-place predicate $U$ that selects elements of the main set $M$ and a distinguished one-place predicate $List=\emptyset$ (a predicate that will select list elements), then we will show how easy it is to prove the following statement about hereditarily finite lists $HW(M)$, which was already proven earlier in \cite{b_osp_pon}, but using a different technique:\\

\textbf{Corollary 3.1:} If $\mathfrak{M}$ - $\Delta_0^p$-model, then $HW(M)$ - is $\Delta_0^p$-set.\\

$\square$ A countable generating family of formulas $F_{List^+}$ has the following view:
\begin{center}
$\varphi_n:\ \&_{i=1}^n(U(x_n)\vee List(x_n)),\ n\in N$
\end{center}

This family of formulas is predicate-separable, all formulas are positive quantifier-free, and the predicate $List$ is included in formulas positively. And how easy it is to check the operator $\Gamma^{\mathfrak{M}}_{List^+}$ is a $\Delta_0^p$-operator. $\blacksquare$\\

Let the signature $\sigma$ have the view: $\sigma=\{c_0,...,c_k,f_1^{(m_1)},...,f_s^{(m_s)},R_1^{(p_1)},...,R_t^{(p_t)}\}$. Consider the model $\mathfrak{N}$ with the basic set of elements $N$ and signatures $\sigma=\{\underline{1},s^{(1)}\}$. The interpretation of the constant $\underline{1}$ will be 1 and $s$-the standard successor function. Further, an entry of the view $\underline{n+1}$ will mean a term of the form $n$-fold application of the function $s$ to $\underline{1}$.\\

\textbf{Corollary 3.2:} The set of quantifier-free formulas of signature $\sigma$ is a $\Delta_0^p$-set.\\

$\square$ The process of constructing auxiliary $\Delta_0^p$-sets using generating families for the corresponding predicates in the $\Delta_0^p$-model $\mathfrak{N}$ is as follows:\\

1) Constants - $F_{Cons^+}$: $\varphi_i: (x_1=\underline{1})\&(x_2=\underline{i})$, $i\in [1,...,k]$\\
2) Variables - $F_{Var^+}$: $\varphi_i: (x_1=\underline{2})\&(x_2=\underline{i})$, $i\in N$\\
3) Function symbols - $F_{Func^+}$: $\varphi_i: (x_1=\underline{3})\&(x_2=\underline{i})$, $i\in [1,...,s]$\\
4) Predicate symbols - $F_{R^+}$: $\varphi_i: (x_1=\underline{4})\&(x_2=\underline{i})$, $i\in [1,...,t]$.\\
5) Terms that are not constants and variables::
\begin{center}
 $F_{Term_1^+}$: $\varphi_i: (x_1=\underline{5})\& Func(x_2)\&_{i=3}^{n_i+2}(Term_1(x_i)\vee Cons(x_i)\vee Var(x_i))$
 \end{center}
6) All set of standard terms:
\begin{center}
 $F_{Term^+}:\ F_{Term_1^+}\cup F_{Cons^+} \cup F_{Var^+}$
 \end{center}
 \textit{Generating family for quantifier-free formulas - $F_{Free^+}$:}\\
1) $\varphi_1(R_i): (x_1=\underline{8})\&R(x_2)\&_{i=3}^{p_i+2}Term(x_i)$\\
2) $\varphi_2(P_i): (x_1=\underline{9})\&P(x_2)\&Term(x_i)$\\
3) $\varphi_3(=): (x_1=\underline{10})\&Term(x_2)\&Term(x_3)$\\
4) $\varphi_4(\&): (x_1=\underline{11})\&Free(x_2)\&Free(x_3)$\\
5) $\varphi_5(\vee): (x_1=\underline{12})\&Free(x_2)\&Free(x_3)$\\
6) $\varphi_6(\to): (x_1=\underline{13})\&Free(x_2)\&Free(x_3)$\\
7) $\varphi_7(\lnot): (x_1=\underline{14})\&Free(x_2)\&Free(x_3)$ $\blacksquare$\\

Expand the signature $\sigma$ to $\sigma'=\sigma\cup\{\in,\subseteq\}$.\\

\textbf{Corollary 3.3:} Set of $\Delta_0$-formulas \cite{b_cond_terms} signature $\sigma'$ is $\Delta_0^p$-set.\\

$\square$ \\
Define a family of $\Delta_0$-formulas $F_{D0^+}$. Just as in Corollary 3.2, we write out generating formulas for terms and formulas, only in our case we also add formulas for the above predicates:\\

8) $\varphi_8(\in): (x_1=\underline{15})\&Term(x_2)\&Term(x_3)$\\
9) $\varphi_9(\subseteq): (x_1=\underline{16})\&Term(x_2)\&Term(x_3)$\\

We also write out generating formulas for $(\exists x_k\in t)\varphi(\overline{x})$, $(\forall x_m\in t)\varphi(\overline{x})$, $(\exists x_t\subseteq t)\varphi(\overline{x})$, $(\forall x_t\subseteq t)\varphi(\overline{x})$:\\

10) $\varphi_{10}(\exists x_k\in t(\overline{x})): (x_1=\underline{17})\&Var(x_2)\&Term(x_3)\&D0(x_4)$\\
11) $\varphi_{11}(\forall x_m\in t(\overline{x})): (x_1=\underline{18})\&Var(x_2)\&Term(x_3)\&D0(x_4)$\\
12) $\varphi_{12}(\exists x_t\subseteq t(\overline{x})): (x_1=\underline{19})\&Var(x_2)\&Term(x_3)\&D0(x_4)$\\
13) $\varphi_{13}(\forall x_t\subseteq t(\overline{x})): (x_1=\underline{20})\&Var(x_2)\&Term(x_3)\&D0(x_4)$ $\blacksquare$\\

\textbf{Corollary 3.4:} The set of conditional terms $\sigma'$ \cite{b_cond_terms} of signature and $\Delta_0^*$-formulas  is $\Delta_0^p$-sets.\\

$\square$ This is where the approach gets more interesting. We need to simultaneously generate both conditional terms and formulas containing these conditional terms. Therefore, we construct two generating families: $F_{TCond^+}$, $F_{FCond^+}$. In addition to generating formulas for standard terms in $F_{TCond^+}$, we add countably many generating formulas for conditional terms:\\
8) $\varphi_{k+8}$: $(x_1=\underline{21})\&(TCond(x_2)\&FCond(x_3))\&...\&(TCond(x_{2k})\&FCond(x_{2k+1}))\&TCond(x_{2k+2})$, $k\in N$.\\
And also the family $F_{FCond^+}$ is defined by the same generating formulas as the family  $F_{D0^+}$, only everywhere the predicates $Term$ must be replaced by $TCond$.\\

$\blacksquare$\\

\textbf{Conclusion.}\\

This work is the starting point for building a methodology for developing fast and reliable software. In this paper, we studied sufficient conditions for the $\Delta_0^p$-operator under which the smallest fixed point remains a $\Delta_0^p-$set. This allows us, based on the main elements, to create new elements and data types. Moreover, for these data types, there are polynomial algorithms for checking whether a certain element belongs to a given data type or not. The question of programming methodology is also of interest: which constructs can be used and which not when creating programs, if we want our programs to be polynomially computable. With the help of the main theorem of our article and the main theorems from the works \cite{b_gonsvir2019},\cite{b_osp_pon} it is already possible to develop logical programming languages, programs in which would be of polynomial computational complexity.


\begin{thebibliography}{9}
\def\bibfont{\small}
\bibfont

\bibitem{b_gonsvir2019}
Goncharov S.S., Sviridenko D.I.: "Logical language of description of polynomial computing."\ \\ Reports of the academy of sciences. Vol 485, No 1 (2019), 11–14

\bibitem{b_alaev1}
P. E. Alaev, V. L. Selivanov, "Polynomial computability of fields of algebraic numbers"\ \\
Dokl. Math., 98:1 (2018), 341–343

\bibitem{b_alaev2}
P. E. Alaev, "Structures Computable in Polynomial Time."\ \\
Algebra and Logic volume 55, pages 421–435(2017)

\bibitem{b_alaev3}
P.E. Alaev, "Existence and Uniqueness of Structures Computable in Polynomial Time"\ \\
Algebra and Logic 55(1), May 2016

\bibitem{b_remmel}
D.Cenzer, J. Remmel, "Polynomial-time versus recursive models"\ \\
Annals of Pure and Applied Logic Volume 54, Issue 1, 26 September 1991, Pages 17-58

\bibitem{b_gonsvir_rec}

Goncharov S.S., Sviridenko D.I.: "Recursive terms in semantic programming"\ \\
Sibirsk. Mat. Zh., 2018, Volume 59, Number 6, Pages 1279–1290

\bibitem{b_cond_terms} 
Goncharov S.S., "Conditional Terms in Semantic Programming."\ \\
In: \textit{Sibirian Mathematical Journal, 2017, V 58, No5, P. 794-800.} 

\bibitem{b_sigma_prog} 
Goncharov S.S., Sviridenko D.I., "$\Sigma$-programming."\ \\
In: \textit{Transl., II.Ser., Am. Math. Soc. 1989. V. 142. P. 101–121.}

\bibitem{b_ges} 
Ershov Yu. L., Goncharov S. S., Sviridenko D. I., "Semantic programming"\ \\
In: \textit{Information processing 86: Proc. IFIP 10-th World Comput. Cong. Dublin.
1986. V. 10. P. 1113–1120.}

\bibitem{b_ershov} 
Ershov Yu. L. "Definability and computability.". 1996

\bibitem{b_osp_pon} 
S. Ospichev, D. Ponomarev, "On the complexity of formulas in semantic programming"\ \\
Sib. Elektron. Mat. Izv. ` , 2018, Volume 15, 987–995

\bibitem{b_harry}
 Harry R. Lewis, Christos Papadimitriou: "Elements of the Theory of Computation". 1998.

\end{thebibliography}
\end{document}